\documentclass[prb,reprint,twocolumn,superscriptaddress,noshowpacs,notitlepage,longbibliography,10pt,citeautoscript,superscriptcitations]{revtex4-2}%

\usepackage{graphicx,bm,times}
\usepackage{pgffor}
\usepackage{amsmath}
\usepackage{amsfonts}
\usepackage{amssymb}
\usepackage{mathtools}
\usepackage{color}
\usepackage{hyperref}
\hypersetup{
	colorlinks = true,
	allcolors = {blue}
}
\setcitestyle{super}

\begin{document}

\title{Resonant interlayer coupling in NbSe$_2$-graphite epitaxial moir{\'e} superlattices}

\author{S.~Mo}
\altaffiliation{These authors contributed equally}
\affiliation{SUPA, School of Physics and Astronomy, University of St Andrews, St Andrews KY16 9SS, United Kingdom}

\author{K.~Kovalenka}
\altaffiliation{These authors contributed equally}
\affiliation{Department of Physics and Astronomy, University of Manchester, Manchester M13 9PL, United Kingdom}

\author{S.~Buchberger}
\affiliation{SUPA, School of Physics and Astronomy, University of St Andrews, St Andrews KY16 9SS, United Kingdom}
\affiliation{Max Planck Institute for Chemical Physics of Solids, Nöthnitzer Strasse 40, Dresden 01187, Germany}

\author{B.K.~Saika}
\affiliation{SUPA, School of Physics and Astronomy, University of St Andrews, St Andrews KY16 9SS, United Kingdom}

\author{A.~Azhar}
\affiliation{Department of Physics and Astronomy, University of Manchester, Manchester M13 9PL, United Kingdom}
\affiliation{Physics Study Program, Faculty of Science and Technology, Syarif Hidayatullah State Islamic University Jakarta, Tangerang Selatan 15412, Indonesia}

\author{A.~Rajan}
\author{A.~Zivanovic}
\affiliation{SUPA, School of Physics and Astronomy, University of St Andrews, St Andrews KY16 9SS, United Kingdom}

\author{Y.-C.~Yao}
\affiliation{SUPA, School of Physics and Astronomy, University of St Andrews, St Andrews KY16 9SS, United Kingdom}
\affiliation{Max Planck Institute for Chemical Physics of Solids, Nöthnitzer Strasse 40, Dresden 01187, Germany}

\author{R.V.~Belosludov}
\affiliation{Institute for Materials Research, Tohoku University, Sendai 980-08577, Japan}

\author{M.D.~Watson}
\affiliation{Diamond Light Source Ltd, Harwell Science and Innovation Campus, Didcot OX11 0DE, United Kingdom}

\author{M.S.~Bahramy}
\email{m.saeed.bahramy@manchester.ac.uk}
\affiliation{Department of Physics and Astronomy, University of Manchester, Manchester M13 9PL, United Kingdom}

\author{P.D.C.~King}
\email{pdk6@st-andrews.ac.uk}
\affiliation{SUPA, School of Physics and Astronomy, University of St Andrews, St Andrews KY16 9SS, United Kingdom}

\date{\today}

\begin{abstract}
Moir{\'e} heterostructures, created by stacking two-dimensional (2D) materials together with a finite lattice mismatch or rotational twist, represent a new frontier of designer quantum materials. Typically, however, this requires the painstaking manual assembly of heterostructures formed from exfoliated materials. Here, we observe clear spectroscopic signatures of moir{\'e} lattice formation in epitaxial heterostructures of monolayer (ML) NbSe$_2$ grown on graphite substrates. Our angle-resolved photoemission measurements and theoretical calculations of the resulting electronic structure reveal moir{\'e} replicas of the graphite $\pi$ states forming pairs of interlocking Dirac cones. Interestingly, these intersect the NbSe$_2$ Fermi surface at the $\mathbf{k}$-space locations where NbSe$_2$'s charge-density wave (CDW) gap is maximal in the bulk. This provides a natural route to understand the lack of CDW enhancement for ML-NbSe$_2$/graphene as compared to a more than four-fold enhancement for NbSe$_2$ on insulating support substrates, and opens new prospects for using moir{\'e} engineering for controlling the collective states of 2D materials.
\end{abstract}

\maketitle

\section*{Introduction}

Super-periodic moir{\'e} potentials in 2D materials have been shown to underpin a plethora of highly tunable interacting electronic states~\cite{balents_superconductivity_2020,andrei_marvels_2021,Castellanos-Gomez2022,mak_semiconductor_2022}, including unconventional superconductors~\cite{cao_unconventional_2018,yankowitz_tuning_2019,lu_superconductors_2019}, correlated insulators~\cite{cao_correlated_2018, chen_evidence_2019, tang_simulation_2020}, and Wigner crystals~\cite{regan_mott_2020,tsui_direct_2024,li_wigner_2024}. Almost all studies of the electronic structure of such moir{\'e} heterostructures to date, however, have been performed using exfoliated materials. While these have been instrumental in helping establish the general phenomenology of 2D moir{\'e} systems, significant materials challenges remain,~\cite{lau_reproducibility_2022} including the achievable levels of twist-angle and strain homogeneity~\cite{uri_mapping_2020,bai_excitons_2020} and the introduction of disorder due to contamination effects induced during the fabrication process~\cite{rosenberger_nano-squeegee_2018,lau_reproducibility_2022}. Despite several pioneering efforts to realise larger-area van der Waals moir{\'e} materials,~\cite{liu_disassembling_2020,jr_macroscopic_2024} this materials complexity has effectively limited the available systems to study to the most stable examples of graphene and the semiconducting transition-metal dichalcogenides (TMDs). Additional approaches for fabricating and controlling moir{\'e} materials are thus strongly desired.

In this respect, ultra-high-vacuum-based epitaxial growth techniques should offer an attractive alternative approach to the formation of large-area and ultra-clean moir{\'e} superstructures from constituent layers hosting different lattice constants. The growth of 2D materials by such methods, however, has traditionally been limited by materials quality, with significant rotational disorder, the formation of small disconnected grains, and premature bilayer formation typically observed~\cite{liu_molecular-beam_2015,walsh_van_2017,Akhil2020,mortelmans_role_2021}. Here, motivated by significant recent enhancements in the quality and uniformity of the 2D layers that can be fabricated~\cite{Akhil2024}, we explore the prospect of creating TMD moir{\'e} heterostructures using molecular-beam epitaxy (MBE). To this end, we synthesise ML-NbSe\textsubscript{2}/graphite van der Waals heterostructures, and study their electronic structure using angle-resolved photoemission spectroscopy (ARPES) and model calculations derived from density functional theory (DFT). Our measurements reveal clear signatures of a well-defined moir{\'e} superstructure being formed, with significant interactions evident between the NbSe$_2$ and graphite layers. Our calculations allow us to identify the origin of the strong interlayer interactions as being due to a resonant coupling between the Fermi surfaces of the graphite and NbSe$_2$ states, and reveal how this can in turn control the collective states of the NbSe$_2$ layer. This opens a new perspective for controlling a pre-existing correlated order using moir{\'e} potentials, while our all-epitaxial approach provides a scaleable platform for the creation of large area moir{\'e} materials.

\section*{Results}

\begin{figure*}
    \centering
    \includegraphics[width=0.95\textwidth]{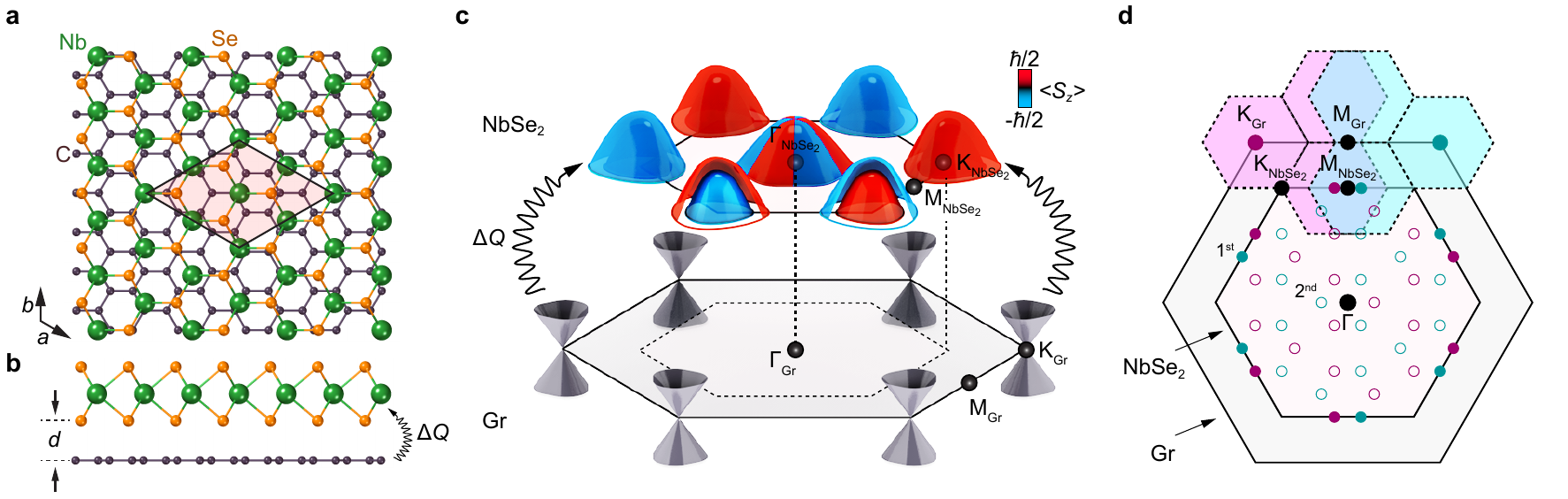}
    \caption{{\bf NbSe$_2$/graphite van der Waals heterostructures.} (a,b) Top (a) and side (b) views of the crystal structure of the ML-NbSe$_2$/graphite heterostructures investigated here. A putative interaction between the NbSe$_2$ and the topmost graphite layer, leading to an interlayer charge transfer, $\Delta{}Q$ is shown. The near coincidence of two unit cells of the NbSe$_2$ to three of the graphite surface layer is indicated by the unit cell shown in (a). (c) Schematic model of the near-$E_\mathrm{F}$ electronic structure of the top-layer graphite and NbSe$_2$ layers. (d) Brillouin zones of the constituent layers, showing the expected location of first-order (solid points) and second-order (open points) moir{\'e} replicas of the K (magenta) and K$^\prime$ (cyan) graphite states, respectively.
    }
    \label{fig:overview}
\end{figure*}

\subsection*{Electronic structure of NbSe$_2$/graphite epitaxial heterostructures}

Fig.~\ref{fig:overview} shows a schematic overview of the NbSe$_2$/graphite heterostructures investigated here. Both NbSe$_2$ and graphite are van der Waals materials, enabling the formation of high-quality epitaxial interfaces between the two (Fig.~\ref{fig:overview}(a,b), see also Supplementary Fig.~1), despite their large lattice constant mismatch of $\approx\!\!{40}$~\%. Unlike most other moir{\'e} systems investigated to date, ML-NbSe$_2$ is itself a correlated material, of strong interest for its CDW state~\cite{Xi2015, Ugeda2016} and as a host of Ising superconductivity~\cite{Xi2016}. The latter arises due to the broken inversion symmetry of the monolayer, combined with an inherently strong spin--orbit coupling. Together, this leads to a spin-polarised electronic structure which is characterised by a locking of the quasiparticle spin to the valley pseudospin~\cite{Xi2016,Bawden2016} (Fig.~\ref{fig:overview}(c)), much like in the famous 2D semiconductors WSe$_2$ and MoS$_2$~\cite{Xiao2012, xu_spin_2014, Riley2014, Suzuki2014}. In fact, NbSe$_2$ can be considered as a significantly hole-doped analogue of MoS$_2$, with a nominal $d^1$ electronic configuration leading to large spin-valley-coupled Fermi surfaces located around the Brillouin zone corners (Fig.~\ref{fig:overview}(c)). Although the Brillouin zone of NbSe$_2$ is much smaller than that of graphite (Fig.~\ref{fig:overview}(d)), for azimuthally-aligned layers as shown in Fig.~\ref{fig:overview}(a), its large Fermi surfaces can be expected to overlap the Dirac states at the K points of the underlying graphite layer. This opens a potential electronic coupling channel between the two layers (Fig.~\ref{fig:overview}(b,c)). If realised, this would raise the exciting prospect to stabilise a rather short-wavelength moir{\'e} heterointerface (Fig.~\ref{fig:overview}(d)) between the strongly lattice-mismatched NbSe$_2$ and graphite layers here; a prospect which we explore in detail below.

\begin{figure*}
    \centering
    \includegraphics[width=\textwidth]{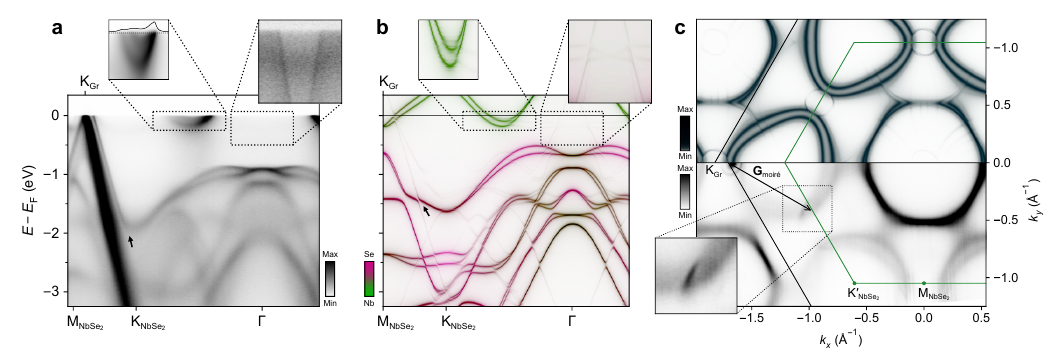}
    \caption{{\bf Measured and calculated electronic structure.} (a) Measured dispersion of the NbSe$_2$/graphite heterostructures along the $\Gamma$-K-M direction, obtained by summing dispersions acquired at $h\nu=70$~eV using circular-left (CL) and circular-right (CR) polarised light. A magnified view of the Nb $d$-orbital metallic bands is shown in the left inset (measured at $h\nu=55$~eV using CL polarisation; the black curve is a momentum distribution curve at the Fermi level). The right inset shows a measurement ($h\nu=55$~eV, linear-vertical polarisation) close to $\Gamma$ with enhanced contrast, showing first-order Nb and second-order graphite moir{\'e} replicas. (b) Corresponding model calculations of the expected dispersion of our heterostructures. The colour shows the atomic character, and the intensity the spectral weight. The inset close to $\Gamma$ is shown with enhanced contrast. (c) Equivalent model calculations (top) and measurements (bottom, $h\nu=70$~eV, CL and CR polarisations) of the Fermi surface. A magnified view of the observed graphite replica is shown in the inset with higher contrast.
    }
    \label{fig:electronic_str}
\end{figure*}

The electronic structure as measured from our fabricated epitaxial heterostructure (see Methods) is shown in Fig.~\ref{fig:electronic_str}(a). We resolve a pair of states crossing the Fermi level along the $\Gamma$-K direction (see left inset). These are nearly degenerate close to $\Gamma$, but become split by around $150$~meV where they cross the Fermi level closer to the zone-corner K point. We assign these as the Nb 4$d$ states, whose band dispersion is in good agreement with our theoretical calculations (Fig.~\ref{fig:electronic_str}(b), left inset). Consistent with previous studies~\cite{Xi2016,Bawden2016,de_la_barrera_tuning_2018}, our calculations (see Methods) indicate that the splitting evident in these states derives from the spin--orbit interaction (see Supplementary Fig.~2), leading to the spin-valley-locked texture shown schematically in Fig.~\ref{fig:overview}(c). This spin splitting has been challenging to observe in electronic structure measurements of single-layer NbSe$_2$ to date~\cite{Dreher2021, Zhang2022}. Here, however, we clearly resolve the band splitting both in our measured dispersions (inset of Fig.~\ref{fig:electronic_str}(a)), and in Fermi surface measurements (bottom panel of Fig.~\ref{fig:electronic_str}(c)), where they contribute a pair of split-off Fermi pockets around each zone-corner K and K$^\prime$ point of the NbSe$_2$ Brillouin zone. Their clear resolution here not only confirms the significant spin--orbit splitting which underpins Ising superconductivity in this system~\cite{Xi2016,de_la_barrera_tuning_2018}, but also points to the high-quality nature of our epitaxial NbSe$_2$ layers. Closer to $\Gamma$, the spin splitting is predicted to become very small, with spin degeneracies formed along the $\Gamma$-M direction (Fig.~\ref{fig:overview}(c) and Fig.~\ref{fig:electronic_str}(c)). Consistent with this, we resolve only a single Fermi pocket centred at the Brillouin zone centre.

At higher binding energies, we observe dispersive states derived from the Se~4$p$ orbitals (Fig.~\ref{fig:electronic_str}(a,b)).~\cite{nakata_anisotropic_2018} Interestingly, we find that these hybridise with the $\pi$ states of the graphite layer where the two intersect close to the boundary of the NbSe$_2$ Brillouin zone (arrow in Fig.~\ref{fig:electronic_str}(a)). This points to a non-negligible interlayer coupling in this system. Consistent with this, we find that the $\pi$ states themselves are split, as compared to measurements of the same bands in a pristine graphite sample (see Supplementary Fig.~3): not only do we observe the expected~\cite{Zhou2006} rather broad (due to $k_z$-dispersion) linearly-dispersing band with its charge neutrality point close to the Fermi level, but also a sharp copy that is shifted to lower binding energy (Fig.~\ref{fig:electronic_str}(a)). The latter contributes a well-defined hole-like Fermi pocket at the zone-corner K point of the graphite Brillouin zone (Fig.~\ref{fig:electronic_str}(c)). It is this latter band that is hybridised with the Se $p$-states from the NbSe$_2$ layer. We thus conclude that the interaction between the NbSe$_2$ layer and the graphite beneath is rather strong, with a significant charge transfer from the topmost graphite layer to the NbSe$_2$ monolayer (Fig.~\ref{fig:overview}(b)) causing the graphite $\pi$ states to become energetically split-off from the bulk-like graphite states below. From a Luttinger analysis (see Supplementary Fig.~3), we estimate that the charge transfer causes a substantial hole doping of the topmost graphite layer of $\approx\!2.5\times10^{13}$ holes/cm$^2$.

\subsection*{Moir{\'e} replica formation}
\begin{figure*}
    \centering
    \includegraphics[width=\textwidth]{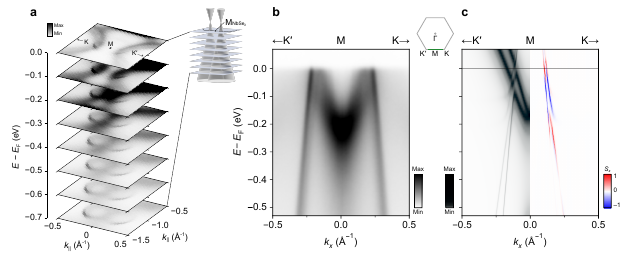}
    \caption{{\bf Moir{\'e} replicas of the graphite $\pi$ states.} (a) Measured constant energy contours ($h\nu=70$~eV, sum of spectra acquired using CL and CR polarisations) in the vicinity of the M point of the NbSe$_2$ Brillouin zone. Clear graphite replicas are visible contributing small pockets at the Fermi level, which grow in area and start overlapping with increasing binding energy below $E_\mathrm{F}$, as shown schematically in the inset. (b) Measured dispersion ($h\nu=55$~eV, CL and CR polarisations) along the K$^\prime$-M-K direction (as indicated in the inset), showing the sharp graphite replica states. (c) Corresponding calculated dispersion along K$^\prime$-M-K, from our supercell calculations. The left-hand side shows the calculations projected onto the NbSe$_2$ layer, while the right shows the spin polarisation of the graphite states when projected onto the C orbitals.
    }
    \label{fig:moire_replicas}
\end{figure*}

Besides this charge transfer and signatures of interlayer hybridisation, the electronic structure features described above would be broadly expected from considering that of individual NbSe$_2$ and graphite layers. We find, however, that there are additional features in the measured electronic structure that can only be obtained for the heterostructure configuration. Most obviously, we observe weak but sharp arc-like features in our measured Fermi surface (Fig.~\ref{fig:electronic_str}(c), see also Supplementary Figs.~4 and 5), intersecting each of the Nb-derived zone-corner Fermi pockets close to the M point of the NbSe$_2$ Brillouin zone. We show a magnified view of one of these in the inset of Fig.~\ref{fig:electronic_str}(c), and constant energy contours in the vicinity of K-M-K$^\prime$ in Fig.~\ref{fig:moire_replicas}(a). From these, it is evident that the observed arcs are in fact one side of a closed trigonally-warped pocket. A clear linear dispersion is evident in our measurements of these features performed along the K-M-K$^\prime$ direction of the NbSe$_2$ Brillouin zone (Fig.~\ref{fig:moire_replicas}(b)), with two cone-like features visible in our constant energy surfaces. These cones start to intersect with increasing energy below the Fermi level. The size and shape of the resulting pockets are, in fact, in excellent agreement with those of the sharp $\pi$ state of the topmost graphite layer (Fig.~\ref{fig:electronic_str}(c) and Supplementary Fig.~4), but shifted in momentum by the moir{\'e} vector defined by the lattice mismatch of the NbSe$_2$ and graphite layers (Fig.~\ref{fig:overview}(d)). We thus attribute the additional sharp features observed here as moir{\'e} replicas of the graphite states~\cite{lu_dirac_2022} coming from neighbouring K and K$^\prime$ points of the original graphite Brillouin zone.

To confirm this assignment, we turn to our DFT-derived model calculations, which explicitly include an interlayer coupling between the NbSe$_2$ and a single graphite layer. We consider a commensurate lattice match between two unit cells of the NbSe$_2$ and three of the graphite. This provides a computationally-tractable system size which is nonetheless quite close to the experimental situation (see Methods). Our calculations in Fig.~\ref{fig:electronic_str}(b,c) and Fig.~\ref{fig:moire_replicas}(c), which are projected onto the orbitals of the NbSe$_2$ layer, show excellent agreement with our experimental measurements. In particular, close to the M point of the NbSe$_2$ Brillouin zone, two trigonally-warped pockets are evident intersecting the Nb-derived Fermi pockets (top panel of Fig.~\ref{fig:electronic_str}(c)). These are derived from replicas of the K and K$^\prime$-valley graphite states, respectively. They are centred exactly at the M point in our calculations due to the enforced commensuration, while they are slightly offset from this in our experiments due to deviations in the real heterostructure from the assumed 2:3 lattice mismatch.

From the comparison with our model calculations, we therefore deduce that the appearance of the graphene-like Dirac cone replicas intersecting the NbSe$_2$ K-barrel Fermi surfaces in our experimental measurements constitute a clear spectroscopic signature of moir{\'e}-assisted interlayer tunnelling. These features arise from momentum-conserving scattering processes in which the mismatch between the reciprocal lattice vectors of the two constituent layers is compensated by the moir{\'e} superlattice. This mechanism is analogous to the generalised Umklapp scattering~\cite{lewandowski2021pairing,wallbank2019excess,li2014thermal} encountered in twisted bilayer graphene, wherein interlayer tunnelling $t(\mathbf{k}_{\text{NbSe}_2},\mathbf{k}_\text{Gr})$ is governed by the selection rule
\begin{equation}
\delta(\mathbf{k}_\text{Gr} + \mathbf{G}_\text{moir\'e} - \mathbf{k}_{\text{NbSe}_2}),
\label{eq:sel_rule}
\end{equation}
with $\mathbf{k}_\text{Gr}$ and $\mathbf{k}_{\text{NbSe}_2}$ denoting the crystal momenta of electrons in the graphene and NbSe$_2$ layers, respectively, and $\mathbf{G}_\text{moir\'e}$ a reciprocal lattice vector of the moir{\'e} potential. In this framework, an electron in a state $|\mathbf{k}_\text{Gr} \rangle$ can tunnel coherently into a state $|\mathbf{k}_{\text{NbSe}_2} \rangle$ provided this condition is met.

For the first-order moir{\'e} harmonics, this interlayer tunnelling gives rise to the arc-like spectral features observed near the M points of the NbSe$_2$ Brillouin zone, with the highest intensities found at the momentum-space locations where the wavefunction overlap $\langle \mathbf{k}_\text{Gr} | \mathbf{k}_{\text{NbSe}_2} \rangle$ is maximal. This spectral weight distribution is entirely consistent with our experimental measurements, where we find the intensity of the replica bands to be strongly peaked where they overlap the Nb-derived Fermi surfaces (see also Supplementary Fig.~5). We note that, if the replica observed experimentally were due to simple final-state effects in our measured photoemission spectra, they would have the same spectral weight variation as for the measured primary graphite states at K$_\mathrm{Gr}$, but simply shifted in momentum. As is typically observed due to the so-called `dark corridor' effect,\cite{Bostwick2007,gierz_illuminating_2011, Polley2019} the spectral intensity of our primary graphite states is strongly peaked within the first graphite Brillouin zone (see also Supplementary Fig.~4), almost the opposite $\mathbf{k}$-dependent variation to that observed of our replica features. This directly indicates that the observed replica features derive from the initial-state electronic structure, and reflect a moir{\'e}-induced band hybridisation between the graphite and the NbSe$_2$-derived states.

Our calculations indicate small but finite hybridisation gaps open where the moir{\'e} replicas cross the primary NbSe$_2$-derived states (Fig.~\ref{fig:moire_replicas}(c)). This leads to momentum-selective hybridisation gaps forming in the Fermi surface, localised to the specific $\mathbf{k}$-points where the moir\'e replicas intersect the NbSe$_2$-derived states (Fig.~\ref{fig:electronic_str}(c)). Such behaviour is indicative of a coherent band anti-crossing process, forming mini-gaps governed by the momentum-matching condition $\mathbf{k}_\mathrm{NbSe_2} = \mathbf{k}_\mathrm{Gr} + \mathbf{G}_\text{moir\'e}$. This leads to a `rim'-like gap structure, reminiscent of the hot spots encountered in CDW phases of transition metal dichalcogenides, including NbSe$_2$~\cite{rahn_gaps_2012,kundu2024charge}, where localised coupling leads to selective Fermi surface reconstruction. It may be captured within a minimal two-level Hamiltonian of the form
\begin{equation}
\widehat{H}(\mathbf{k}) = 
\begin{pmatrix}
\epsilon_{\mathrm{NbSe_2}}(\mathbf{k}) & t(\mathbf{k},\mathbf{k} + \mathbf{G}_\text{moir{\'e}}) \\
t^*(\mathbf{k},\mathbf{k} + \mathbf{G}_\text{moir\'e}) & \epsilon_{\mathrm{Gr}}(\mathbf{k} + \mathbf{G}_\text{moir\'e})
\end{pmatrix},
\label{eq:coupling}
\end{equation}
from which the local hybridisation gap follows as $\Delta_{\text{moir{\'e}}}(\mathbf{k}) = 2|t(\mathbf{k},\mathbf{k} + \mathbf{G}_\text{moir{\'e}})|$. 

While it is difficult to directly resolve the gap in our measured dispersions due to the small associated energy scales, we note that the underlying Nb-derived K-barrel Fermi surfaces observed experimentally are more warped than would be expected for an isolated monolayer of NbSe$_2$.~\cite{nakata_anisotropic_2018} In particular, the spin splitting almost vanishes along the M-K direction. This is suggestive of the formation of momentum-selective hybridisation gaps by the moir{\'e} interaction, that open where the primary NbSe$_2$ and graphite replica Fermi surfaces cross each other. This motivates future studies of aligned NbSe$_2$/graphite samples by high-resolution quasiparticle-interference imaging, where the corresponding gap openings at the Fermi surface should be clearly resolvable.~\cite{naritsuka_superconductivity_2025} Interestingly, as shown in Fig.\ref{fig:moire_replicas}(c), our calculations reveal a finite spin polarisation emerging in the graphite bands near the hybridisation gaps. This phenomenon can be directly attributed to the off-diagonal tunnelling elements in Eq.~(\ref{eq:coupling}), which mediate spin transfer from the strongly spin-polarised NbSe$2$ states, $\epsilon_{\mathrm{NbSe_2}}$, to the nominally spin-degenerate states in graphite, $\epsilon_{\mathrm{Gr}}$. The resulting spin texture in the graphite-derived bands highlights that the moir{\'e} interaction can provide a route to proximity-induced spin--orbit coupling; something that is being actively sought for spintronic functionality in graphene-based heterostructures~\cite{avsar_spinorbit_2014,island_spinorbit-driven_2019}.

As well as these `first-order' replicas of the primary graphite states, our calculations (Fig.~\ref{fig:electronic_str}(b,c)) reveal a rich hierarchy of additional replica features. Near the $\Gamma$ point, a rather flat feature is evident in the calculated band dispersions along the $\Gamma$-K direction, close to the band bottom of the Nb-derived conduction bands (Fig.~\ref{fig:electronic_str}(b)). This results from a moir{\'e}-induced replica of the primary NbSe$_2$-derived states (see also the Nb-derived $\Gamma$- and K-barrel replicas evident in our Fermi surface calculations, Fig.~\ref{fig:electronic_str}(c)). Moreover, weak spectral weight from steep graphite-like band dispersions is also visible centred around the $\Gamma$ point (see right inset of Fig.~\ref{fig:electronic_str}(b)). This can be assigned as second-order moir{\'e} replicas of the graphite states (Fig.~\ref{fig:overview}(d)). At second and higher orders, the same selection rule as in Eq.~(\ref{eq:sel_rule}) applies, but the tunnelling amplitude is significantly diminished. The second-order graphite replicas therefore only yield faint features in our calculations, e.g., near the $\Gamma$ point. Nonetheless, while extremely weak, finite signatures of both of the above features can be observed in our experimental measurements (inset close to $\Gamma$ in Fig.~\ref{fig:electronic_str}(a)). While the graphite states are again slightly displaced in momentum from those in our calculations due to the enforced commensurability in the latter, they are otherwise in excellent qualitative agreement.

\section*{Discussion}
The resulting moir{\'e} heterostructure formation thus fundamentally reshapes the low-energy electronic structure of this system. This can be expected to have a direct impact on the collective states which NbSe$_2$ hosts. Indeed, recent scanning tunnelling microscopy quasiparticle interference measurements from rotated structural domains of NbSe$_2$/graphene have shown evidence for momentum-dependent modulations of the superconducting gap arising due to the substrate-epilayer moir{\'e} lattice~\cite{naritsuka_superconductivity_2025}. While for a different azimuthal alignment in the measurements shown here, our direct observation of moir{\'e} replicas in the electronic structure of NbSe$_2$/graphite heterostructures supports their importance for understanding superconductivity in ML-NbSe$_2$, and motivates further studies investigating their detailed dependence on interlayer twist angle.

\begin{figure}
    \centering
    \includegraphics[width=\columnwidth]{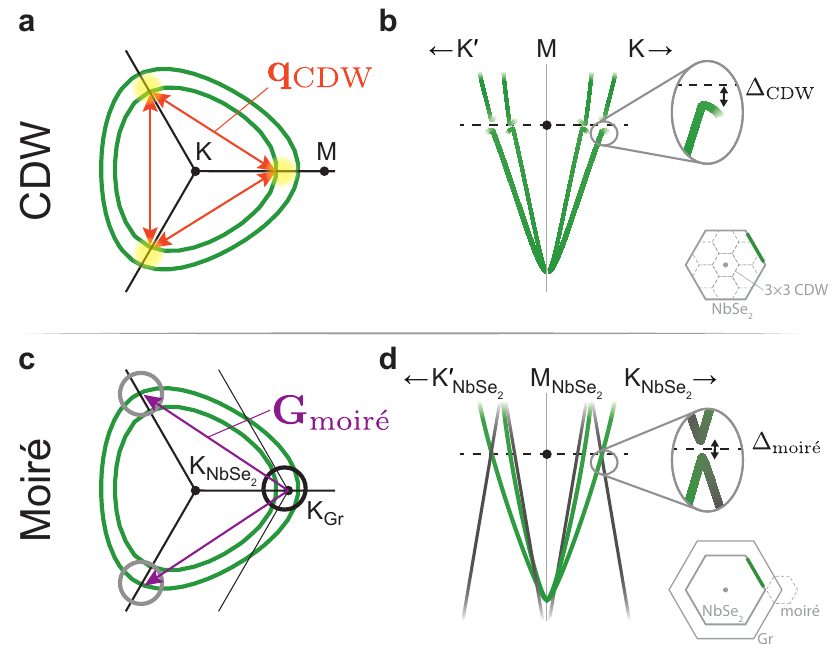}
    \caption{{\bf Impact of moir{\'e} replicas on the CDW order in NbSe$_2$.} (a) Schematic K-barrel Fermi surfaces of NbSe$_2$, showing the wavevectors of the $3\times3$ CDW, $\mathbf{q}_\mathrm{CDW}$ and (b) the corresponding opening of CDW gaps at the Fermi level, with an energy scale $\Delta_{\mathrm{CDW}}$. (c) Corresponding schematic K-barrel Fermi surfaces for ML-NbSe$_2$/graphite heterostructures, showing in addition the primary graphite states (black circle) and their moir{\'e} replicas (grey circles), shifted by $\mathbf{G}_\text{moiré}$. (d) The intersection of the NbSe$_2$ states and moir{\'e} replicas opens hybridisation gaps ($\Delta_\text{moiré}$) at the same $\mathbf{k}$-space locations where the CDW gap is maximal ({\it cf.} (b)). The insets at the bottom right in (b) and (d) show the corresponding Brillouin zones, showing also the $(3\times3)$ CDW and moir{\'e} zones, respectively.}
    \label{fig:summary}
\end{figure}
Moreover, we note that the $(3\times3)$ CDW instability in NbSe$_2$ is known to open its dominant energy gaps at the corners of the K-barrel Fermi surfaces (Fig.~\ref{fig:summary}(a,b)), with the largest gap opening on the smaller pocket.~\cite{rahn_gaps_2012} Experimentally, we find that the graphite replicas cross the Nb-derived Fermi pockets precisely at this momentum-space location (Fig.~\ref{fig:moire_replicas}(a), Supplementary Figs.~4 and 5), as shown schematically in Fig.~\ref{fig:summary}(c). We estimate the energy scale of the resulting moir{\'e}-induced hybridisation gaps at the Fermi level (Fig.\ref{fig:summary}(d)) as $\Delta_\text{moir{\'e}}\approx10$~meV from our full DFT supercell calculations. This is larger than the largest reported CDW gaps in the bulk~\cite{rahn_gaps_2012}, and so can be expected to directly suppress the thermally-sensitive CDW instability. Indeed, such a competition may provide a microscopic mechanism to rationalise the controversial thickness-dependent CDW evolution in NbSe$_2$: For ML-NbSe$_2$/graphene heterostructures, where this competition can be expected, the CDW has been shown to become stable only at or below the bulk transition temperature.~\cite{Ugeda2016} In contrast, for samples supported on insulating substrates, where there are no states available to open hybridisation gaps at the Fermi level from moir{\'e} coupling at the CDW wavevector, the CDW is known to be significantly strengthened in the monolayer limit.~\cite{Xi2015,lin_patterns_2020}

Our results therefore show that, beyond the well-established ability of moir{\'e} lattice formation to stabilise a plethora of correlated states in 2D materials, it may also act to {\it destabilise} collective states that are inherent to the constituent materials. As many of the 2D materials which host such states are intrinsically air sensitive, we expect that our epitaxial moir{\'e} approach will provide a powerful materials platform for investigating this.

{\small
\section*{Methods}
\noindent{\bf{Molecular beam epitaxy:}}
The samples were grown on natural graphite substrates, which were secured to Ta foil chips using high-temperature carbon paste. Before growth, the paste was cured at approximately 300$^{\circ}$C, and the substrates were then degassed at 600$^{\circ}$C for approximately 2 hours at a pressure of $\sim\!10^{-7}$~mbar. The graphite substrates were exfoliated just before being loaded into the vacuum system, where they were further degassed in the load lock at $200^\circ$C for 10 hours. Immediately before the thin film growth, they were finally annealed at $800^\circ$C for $\sim30$ minutes.
 
The NbSe\textsubscript{2} films were grown at a nominal growth temperature of 700$^\circ$C, measured from a thermocouple behind the substrate. Nb was evaporated from a \textit{FOCUS EFM} electron beam evaporator, maintaining a supply of 2.0~nA measured by the integrated flux monitor. Se was evaporated from a valved cracker cell with temperatures of 163 and 500$^\circ$C for the tank and cracking zone, respectively, resulting in a beam equivalent pressure of $2\times10^{-7}$~mbar. A 3N5 pure Nb rod and 5N pure Se granules were used as source materials. The growth was monitored by reflection high-energy electron diffraction, confirming a high monolayer coverage after the growth time of 2~h~10~min, as shown in Supplementary Fig.~1. The growth was ended by cutting off the Nb supply and the sample was cooled down under the Se pressure, which was cut off at $300^\circ$C.

\

\noindent{\bf{Angle-resolved photoemission:}} Micro-ARPES measurements were performed at the nano-ARPES end station of the I05 beamline at Diamond Light Source, using a capillary focusing optic with a beam spot size on the sample of 4--5~$\mu$m. The samples were transferred from the growth system to the beamline end station using a vacuum suitcase. The samples were cooled to approximately 25~K and measured using a Scienta DA30 electron analyser. The analyser slit direction is represented as the $k_x$ direction in the data. The photon energies and polarisations used are indicated in the figure captions. 

\

\noindent{\bf{Theoretical calculations:}}
To construct a realistic model of the electronic structure, we first performed DFT calculations of a NbSe$_2$/graphene heterostructure employing the Perdew–Burke–Ernzerhof exchange–correlation functional~\cite{pbe}, as implemented in the VASP package~\cite{Kresse_1996,Kresse_1999}. The simulations were carried out on a trigonal supercell comprising $2\times2$ unit cells of NbSe$_2$ and $3\times3$ unit cells of graphene. To eliminate lattice mismatch within the supercell, the in-plane lattice constant of graphene was isotropically compressed from its experimental value of 2.460~\AA~\cite{yang2018structure} to 2.294~\AA, while the NbSe$_2$ lattice constant was maintained at its experimental value of 3.442~\AA~\cite{MEERSCHAUT20011721}. A vacuum layer of 18~\AA~was introduced along the crystallographic $c$-axis to prevent spurious interactions between periodic images. The atomic positions were fully relaxed until the residual forces on all atoms fell below 0.001 eV/~\AA, using a kinetic energy cut-off of 400 eV for the plane waves included in the basis set. Spin--orbit coupling was explicitly included, and the Brillouin zone was sampled using a $12\times12\times1$ Monkhorst–Pack $k$-mesh. From the converged band structure, we extracted the energy offset between the graphene Dirac point and the Fermi level.

We then performed analogous DFT calculations for isolated monolayers of NbSe$_2$ and graphene, using the same unit cell geometries but without structural relaxation. The resulting Hamiltonians were projected onto Wannier functions~\cite{Mostofi2014} using Nb-$4d$, Se-$4p$, and C-$2p_z$ orbitals as projection centres, yielding 88-band ($\widehat{H}_{\text{NbSe}_2}$) and 18-band ($\widehat{H}_{\text{Gr}}$) tight-binding models, respectively. To align the energy scales of the two subsystems, appropriate onsite energy corrections were applied based on the energy shift determined in the full heterostructure calculation. The total Hamiltonian of the heterostructure was then constructed as $\widehat{H} = \widehat{H}_{\text{NbSe}_2} + \widehat{H}_{\text{Gr}} + \widehat{H}_{\text{int}}$, where the interlayer coupling term $\widehat{H}_{\text{int}}$ was derived by introducing orbital-dependent hopping interactions between C and Nb atoms as well as C and Se atoms. Specifically, $\widehat{H}_{\text{int}}$ includes coupling terms between all C-$2p_z$ orbitals and in-plane $\{\text{Nb-}4d_{xy}, \text{Nb-}4d_{x^2-y^2}\}$ and out-of-plane Nb-$4d_{z^2}$ orbitals, as well as between all C-$2p_z$ orbitals and in-plane $\{\text{Se-}4p_x,\text{Se-}4p_y\}$ and out-of-plane Se-$4p_z$ orbitals of the Se layer adjacent to graphene. The interlayer interaction $t_{ij}$ between $i=\text{C}$ and $j=\{\text{Nb, Se}\}$ was modelled as an exponentially decaying function $t_{ij} = \lambda_j\exp(-r^{||}_{ij}/\xi)$, where $r^{||}_\text{ij}$ is the in-pane distance between sites $i$ and $j$. To obtain the intersite hopping matrix element, this interlayer coupling term modulates the orbital overlaps, which were fully accounted for in order to filter the active tunnelling channels in the hybridisation spectrum based on the orbital composition and symmetry of the constituent Bloch wavefunctions. For the sake of simplicity, we treat the modulating scalar factor for in-plane and out-of-plane orbitals on an equal footing: the modulation parameter $\xi=3$~\AA~, and $\lambda_\text{Nb}=0.30$~eV and $\lambda_\text{Se}=0.36$~eV. Finally, the unfolded electronic band structures and constant-energy maps were computed using a band unfolding procedure \cite{PhysRevLett.104.216401}, enabling direct comparison with the experimental ARPES spectra.

\section*{Acknowledgements}
We thank Martin McLaren and Rick Davitt for technical assistance. We gratefully acknowledge support from the UK Engineering and Physical Sciences Research Council (Grant No.~EP/X015556/1). We thank Diamond Light Source for access to Beamline I05 (Proposal SI36192), which contributed to the results presented here. We also gratefully acknowledge the Research Infrastructures at the Center for Computational Materials Science at the Institute for Materials Research for allocations on the MASAMUNE-IMR supercomputer system (Project No. 202112-SCKXX-0510) and MAHAMERU BRIN HPC facility under the National Research and Innovation Agency of Indonesia.  M.S.B. acknowledges support from Leverhulme Trust (Grant No. RPG-2023-253). K. K. was supported by the CDT in Graphene NOWNANO. A.A. had support from the Indonesia Endowment Fund for Education (LPDP), NIB/202209223311735. R.V.B. and M.S.B. are grateful to E-IMR center at the Institute for Materials Research, Tohoku University, for continuous support. For the purpose of open access, the authors have applied a Creative Commons Attribution (CC BY) licence to any Author Accepted Manuscript version arising. The research data supporting this publication can be accessed at [[DOI TO BE INSERTED]].

}

% \clearpage

\foreach \x in {1,2,3}
{%
\begin{figure*}[!ht]
  \centering
  \includegraphics[
  width=\textwidth,
  page=\x,
  trim=1.5cm 1cm 1.5cm 1.5cm,
  clip
]{supp.pdf}
\end{figure*}
}

\end{document}